\documentclass[preprint,preprintnumbers,amsmath,amssymb]{revtex4}
\usepackage{graphicx}% Include figure files
\usepackage{bm}% bold math

\begin{document}

\title{Quantum renormalization of entanglement in an antisymmetric anisotropic and bond-alternating spin system}

\author{Xiang Hao}
\altaffiliation{Corresponding author} \email{110523007@suda.edu.cn}

\affiliation{Department of Physics, School of Mathematics and
Physics, Suzhou University of Science and Technology, Suzhou,
Jiangsu 215011, People's Republic of China}

\begin{abstract}

The quantum renormalization group method is applied to study the
quantum criticality and entanglement entropy of the ground state of
the Ising chain in the presence of antisymmetric anisotropic
couplings and alternating exchange interactions. The quantum phase
transitions can be characterized by the discontinuity in the second
derivative of the energy of renormalized ground state. The phase
diagram is obtained by the critical boundary line. The first
derivative of entanglement entropy also diverges at the same
critical points after enough iterations of the renormalization of
coupling constants. The antisymmetric anisotropy and alternating
interaction can enhance the renormalized entanglement via the
creation of quantum fluctuations. The scaling behavior of the
derivative of the entropy around the critical points manifest the
logarithm dependence on the size of the spin system.

\end{abstract}

\maketitle

\section{Introduction}

The resource of entanglement can play an important role in the
quantum information processing such as quantum teleportation and
algorithms for quantum computation \cite{Divin98,Nielsen00}. The
entanglement is quantum correlation which can describe the property
of the ground state for various spin systems \cite{Vedral08}. It has
been justified that there exist the close relations between the
entanglement and the quantum phase transitions in condensed matter
physics
\cite{Osborne02,Bose02,Gu04,Mosseri04,Lidar04,Verstraete04,Anfossi07}.
The quantum critical phenomena happen at zero temperature when the
spin couplings and external fields are varied in the vicinity of
critical points \cite{Sachdev00}. As is known, the quantum phase
transitions can be represented by the discontinuities in the
derivatives of some order parameters. However, for no \emph{a
priori} knowledge of suitable order parameters, some special
physical quantities can be used to universally identify the quantum
criticality at the ground state \cite{Zanardi07, Yang07}. The
ground-state energy is considered as one characterizing the quantum
phase transitions for some spin systems \cite{Lou04, Lou05}. So far
some measures of quantum entanglement have been found out to be
relevant to the quantum criticality for many spin systems
\cite{Osterloh02,Vidal03,Lou06}. During the theoretical
investigation of low-dimensional spin systems, the numerical methods
of density-matrix renormalization group and exact diagonalization
have received much attention \cite{Schollwock05,Legeza03,Li09}. In
many materials, the magnetic properties were studied by the spin
models with the antisymmetric anisotropic interaction such as
Dzyaloshinskii-Moriya interaction \cite{Dender97,Fulde95}. Recently,
the quantum renormalization group approach has provided a convenient
way to analyze the quantum phase transitions for such systems
\cite{Kargarian07,Kargarian08,Kargarian082,Kargarian09}. By this
method, the renormalized ground states can be described in terms of
matrix product states \cite{Verstraete05}. The quantum phase diagram
can be approximately obtained by the enough iterations of the
renormalization of couplings and external fields. Meanwhile, rich
quantum criticality can also be embodied by the bond-alternating
spin models \cite{Sierra07}. These respects motivate us to
investigate the effects of antisymmetric anisotropy and alternating
spin interactions on the entanglement and quantum phase transitions.

The paper is organized as follows. In section II, the quantum
renormalization group method is introduced and used for the Ising
chain with antisymmetric anisotropy and alternating exchange
interactions. The renormalized ground state can be obtained
analytically. In section III, the quantum phase transitions are
approximately characterized by the energy of the renormalized ground
state. The boundary line of critical points determines two different
phases. The effects of antisymmetric anisotropic interactions and
alternating ones on the entanglement entropy are investigated. The
divergence of the derivative of entropy occurs at the same critical
points. The finite size scaling behavior of the entanglement are
also studied. A simple discussion concludes the paper.

\section{Renormalization of the spin model}

The implementation of quantum renormalization group is the
perturbation method to reduce the degrees of freedom of the spin
model. In the Kadanoff's block approach \cite{Sierra97}, the
original Hamiltonian of the model can be divided into the block
Hamiltonian and the interacting Hamiltonian. The block Hamiltonian
is the sum of some commuting items which act on different blocks.
The effective Hamiltonian can be obtained by mapping the original
Hamiltonian to the low energy subspace via the projector consisting
of some lowest energy eigenstates of the block Hamiltonian.

The Hamiltonian of the Ising chain with the antisymmetric
anisotropic and alternating interactions can be written by
\begin{equation}
H=J\sum_{i=1}^{N}[1-(-1)^i\lambda]S_{i}^{z}S_{i+1}^{z}+\vec{D}\cdot(\vec{S}_{i}\times
\vec{S}_{i+1}).
\end{equation}
The periodic boundary condition of $N+1=1$ is considered. Here
$\vec{S}_{i}=(S_{i}^{x},S_{i}^{y},S_{i}^{z})$ denotes the Pauli
operator acting on the $i$-th spin and
$S_{i}^{z}|\uparrow(\downarrow)\rangle=\pm \frac 12
|\uparrow(\downarrow)\rangle$. The case of $J>0$ represents the
nearest-neighboring antiferromagnetic coupling and $0\leq\lambda
\leq 1$ describes the relative strength of alternating exchange
coupling. The antisymmetric anisotropic interaction is described by
the vector of the Dzyaloshinskii-Moriya interaction,
$\vec{D}=(D_{x},D_{y},D_{z})$, arising from the spin-orbit coupling.
In the following analysis, the case of $D_{x}=D_{y}=0,D_{z}=D$ is
taken into account.

According to the renormalization group method, the original
Hamiltonian is divided into two parts
\begin{equation}
H=H^{0}+H^{I}=\sum_{k=1}^{L}h_{k}^{0}+h_{k,k+1}^{I}.
\end{equation}
The part of $H^{0}=\sum_{k=1}^{L}h_{k}^{0}$ is the block Hamiltonian
where $L=N/3$ represents the number of blocks and each item has
three sites,
$h_{k}^{0}=J\sum_{l=1}^{2}[1+(-1)^{k+l}\lambda]S_{l,k}^{z}S_{l+1,k}^{z}+D(S_{l,k}^{x}S_{l+1,k}^{y}-S_{l,k}^{y}S_{l+1,k}^{x})$.
Here $S_{l,k}^{\alpha}(\alpha=x,y,z)$ denotes the $\alpha$-component
pauli operator on the $l$-th spin of the $k$-th block. The other
part of $H^{I}$ represents the interacting Hamiltonian between two
neighboring blocks, $H^{I}=\sum_{k=1}^{L}h_{k,k+1}^{I}$ where
$h_{k,k+1}^{I}=J\{[1-(-1)^{k}\lambda]S_{3,k}^{z}S_{1,k+1}^{z}+D(S_{3,k}^{x}S_{1,k+1}^{y}-S_{3,k}^{y}S_{1,k+1}^{x})
\}$. We need the degenerate lowest energy eigenstates
$|\psi_{k}^{\pm}\rangle$ of $h_{k}^{0}$ to construct the projector
$P_{k}=|\Uparrow_{k}\rangle \langle
\psi_{k}^{+}|+|\Downarrow_{k}\rangle \langle \psi_{k}^{-}|$ where
$|\Uparrow(\Downarrow)\rangle$ is the renamed basis in the effective
Hilbert space for each block. When $k$ is odd, the analytic
expression of $|\psi_{k}^{\pm}\rangle$ can be given in the product
Hilbert space of a block
\begin{align}
|\psi_{k}^{+}\rangle&=a|\uparrow \uparrow
\downarrow\rangle+ib|\uparrow \downarrow \uparrow
\rangle+c|\downarrow \uparrow \uparrow \rangle \nonumber \\
|\psi_{k}^{-}\rangle&=a|\downarrow \downarrow
\uparrow\rangle-ib|\downarrow \uparrow \downarrow \rangle+c|\uparrow
\downarrow \downarrow \rangle.
\end{align}
Here the real parameters of $a,b,c$ satisfy the condition of the
normalization where $a=bD/(\lambda-2\epsilon_{0}),b=1/\sqrt
{1+2D^2(4\epsilon_{0}^{2}+\lambda^2)/(4\epsilon_{0}^{2}-\lambda^2)^2}$.
Here $\epsilon_{0}$ is the scale of the lowest energy by the
parameter $J$ and calculated by the minimal root of the equation
$8\epsilon_{0}^{3}+4\epsilon_{0}^{2}-2(\lambda^2+2D^2)\epsilon_{0}-\lambda^2=0$.
For the simplest example of $\lambda=0$,
$\epsilon_{0}=-(1+\sqrt{1+8D^2})/4$. When $k$ is even, the
degenerate lowest energy eigenstates have the similar expression in
Eq. (3) and the real parameters
$b_{even}=b,a_{even}=-c,c_{even}=-a$. The effective Hamiltonian
mapped onto the low energy subspace can be expressed by
\begin{equation}
H_{eff}=(\prod_{k=1}^{L}P_{k})H(\prod_{k=1}^{L}P_{k}^{\dag}).
\end{equation}
By the effective spin operators $\vec{\tau}_k$ for the $k$-th block,
the definite form in the effective Hilbert space can be written by
$H_{eff}=J^{\prime}\sum_{k=1}^{L}[1-(-1)^k\lambda]\tau_{k}^{z}\tau_{k+1}^{z}+D^{\prime}(\tau_{k}^{x}\tau_{k+1}^{y}-\tau_{k}^{y}\tau_{k+1}^{x})$
where the operator $\tau^{z}=\frac 12(|\Uparrow\rangle \langle
\Uparrow|-|\Downarrow\rangle \langle \Downarrow|)$. For the case of
$k=$ odd, the relations between the spin operator
$\vec{S}_{l,k}=(S_{l,k}^{x},S_{l,k}^{y},S_{l,k}^{z})$ and the
effective one
$\vec{\tau}_{k}=(\tau_{k}^{x},\tau_{k}^{y},\tau_{k}^{z})$ are given
by
\begin{align}
P_{k}\vec{S}_{1,k}P_{k}^{\dag}&=\left(
2ab\tau_{k}^{y},-2ab\tau_{k}^x,(1-2c^2)\tau_{k}^{z} \right)\nonumber
\\
P_{k}\vec{S}_{2,k}P_{k}^{\dag}&=\left(
2ac\tau_{k}^{y},2ac\tau_{k}^x,(1-2b^2)\tau_{k}^{z} \right)\nonumber
\\
P_{k}\vec{S}_{3,k}P_{k}^{\dag}&=\left(
2bc\tau_{k}^{y},-2bc\tau_{k}^x,(1-2a^2)\tau_{k}^{z} \right).
\end{align}
The renormalization of the couplings satisfy the equation
\begin{equation}
J^{\prime}=J(1-2a^2)(1-2c^2),D^{\prime}=-\frac
{4Dab^2c}{(1-2a^2)(1-2c^2)}.
\end{equation}
In general, after the $n$-th renormalization group iteration step,
the chain of $3^{n+1}$ sites can be represented by the effective
three-site Hamiltonian with the renormalized couplings. According to
the renormalization of ground state given by the expression of Eq.
(3), the quantum phase transitions and the entanglement properties
can be studied.

\section{Phase diagram and entanglement entropy}

The quantum phase transitions at the ground state are driven by the
quantum fluctuations. It is the key to select the suitable order
parameters. To obtain the phase diagram, we choose the ground-state
energy as one universal measure. From the renormalization equation
of couplings in Eq. (7), it is seen that the quantum phase
transitions will happen by the change of the antisymmetric
anisotropic interaction $D$ and relative strength of alternating
spin exchange $\lambda$. For a certain value of $\lambda$, the
obvious discontinuity in the second derivative of ground-state
energy $\frac {\partial^2\epsilon_0}{\partial D^2}$ can be shown at
the quantum critical point $D_{c}$ in Fig. 1(a). It means the
feature of the second-order quantum phase transition. With the
change of the alternating interaction $\lambda$, the phase diagram
is obtained in Fig. 1(b). It is shown that two different phases are
divided by the critical points $D_{c}\simeq \sqrt{1-\lambda^2}$.

It is interesting to investigate the entanglement properties of the
spin system in the two different phases. Using the renormalization
group method, we calculate the entanglement between some degrees of
freedom and the rest of the system. The renormalization of
entanglement in the large enough iteration step can be used to
estimate the large size behavior of the spin system. As one
successful measure of entanglement, the block entropy can be
implemented to describe the global property of the entanglement. The
density matrix of the renormalized ground state is
$\rho_{0}=|\psi_{0}\rangle \langle \psi_{0}|$ where
$|\psi_{0}\rangle=|\psi_{0}^{+}\rangle$ or $|\psi_{0}^{-}\rangle$ is
expressed by Eq. (3). Therefore, the reduced density matrix for the
middle site can be obtained by tracing over the rest sites
\begin{equation}
\rho=(1-b^2)|\Uparrow\rangle \langle \Uparrow|+b^2|\Downarrow\rangle
\langle \Downarrow|.
\end{equation}
The entropy $E$ for the middle site can represent the entanglement
between the degrees of freedom of the middle site and ones of the
rest sites. The expression of $E$ is given by
\begin{equation}
E=-(1-b^2)\log_2(1-b^2)-b^2\log_2(b^2).
\end{equation}
In the first step of renormalization group, the chain of $3^2$ sites
can be described by the effective Hamiltonian $H_{eff}$. The change
of entanglement entropy $E$ is depicted as a function of the
antisymmetric anisotropic coupling $D$ and alternating exchange
interaction $\lambda$ in Fig. 2. It is shown that the values of $E$
are increased with $D$ and $\lambda$. The Dzyaloshinskii-Moriya
interaction in the $XY$ plane can induce the planar quantum
fluctuations which enhance the quantum correlation of the ground
state. The existence of alternating couplings can spoil the form of
the N\'{e}el ordered state which has no entanglement. The values of
the entanglement tend to one maximum when $\lambda$ is close to one.

To analyze the connection of the entanglement to the quantum phase
transition, we need study the scaling property. By means of
increasing the iteration step, the entropy $E$ can be calculated for
the large-size spin system. From Fig. 3, it is seen that the rate of
the change of entanglement entropy will be increased as the size
becomes large through the iteration step. In the phase $I$ for the
case of $D<D_{c}$, the entropy is always less than one. The maximal
entanglement entropy of $E=1$ occurs in the phase $II$ for the case
of $D>D_{c}$. After the large enough iteration step, the appearance
of nonanalytic behavior of the entanglement entropy verifies the
feature of the quantum phase transition. In this spin system, the
measure of entanglement entropy can be applied to estimate the
quantum criticality.

Furthermore, it is also necessary to evaluate the singularity of the
derivative of the renormalized entanglement entropy. As depicted in
Fig. 4(a), the drop of the first derivative of entropy will become
more pronounced with the increase of the spin sites. The minimal
values of $-\frac {\partial E}{\partial D}$ at some points $D=D_{m}$
are decreased rapidly with the size. The points of $D_{m}$ for a
large iteration step are closely near to the critical ones $D_{c}$.
The scaling of the points $D_{m}$ is also shown in Fig. 4(b). The
logarithm dependence on the sites $N$ is numerically given by
$D_{m}=D_{c}-N^{-\gamma}$ where the parameter
$\gamma=\gamma(\lambda)$ is increased with the decrease of the
alternating coupling $\lambda$. Through the numerical fit,
$\gamma(1)\approx 0.46$ and $\gamma(0.5)\approx 0.52$. This respect
manifests that the divergence of the derivative of the entropy
appears more rapidly for the smaller values of alternating
interaction $\lambda$ in the condition of the same size.

\section{Discussion}

The entanglement property and quantum phase transitions in the Ising
chain with the Dzyaloshinskii-Moriya interaction and alternating
coupling are studied by the method of quantum renormalization group.
Based on the analytic renormalization equation, the renormalized
ground state can be obtained. The ground-state energy is considered
as one universal measure of quantum phase transitions. The
nonanalytic behavior of the second derivative of the renormalized
ground-state energy can justify the feature of second-order phase
transition. The critical boundary line divides the ground state into
two different phases. It is demonstrated that the antisymmetric
anisotropy and alternating couplings can help for the increase of
the entanglement entropy. The discontinuities in the entanglement
entropy happen at the critical points which are same to those
provided via the order parameter of the ground-state energy. With
the decrease of the alternating coupling, the divergence of the
derivative of the entropy appears more rapidly in the vicinity of
the quantum criticality. In this one-dimensional spin system, the
renormalized entanglement entropy is the efficient measure of the
quantum phase transition.

\section{Acknowledgement}

X.H. was supported by the Research Program of Natural Science for
Colleges and Universities in Jiangsu Province Grant No. 09KJB140009
and the National Natural Science Foundation Grant No. 10904104.

\newpage
{\Large Figure caption}

Figure 1

(a). The second derivative of the energy of the renormalized ground
state diverges at the critical point $D_{c}\simeq 0.85$ in the
iteration step $n=9$ when $\lambda=0.5$. (b). The dash line denotes
the critical boundary characterizing two different phases $I$ and
$II$.

Figure 2

The renormalization of the entanglement entropy $E$ is plotted as a
function of the antisymmetric anisotropy $D$ and alternating
coupling $\lambda$ in the iteration step $n=1$.

Figure 3

The evolution of the renormalized entanglement entropy as the size
is shown for $\lambda=0.5$. The square data denote the case of the
step $n=1$, the circle ones represent the case of $n=5$ and the
triangles are the data for the case of $n=9$.

Figure 4

(a). The negative values of the first derivative of the entropy are
calculated when $\lambda=0.5$. The dot line describes the case of
iteration step $n=1$, the dash line denotes the result of $n=5$ and
the solid one represent the case of $n=9$. (b). The data
characterize the finite size scaling property of the singular points
$D_{m}$ where the values of $-\frac {\partial E}{\partial D}$ are
minimal. The squares represent the case of $\lambda=0.5$ and the
circles denote the case of $\lambda=1$.

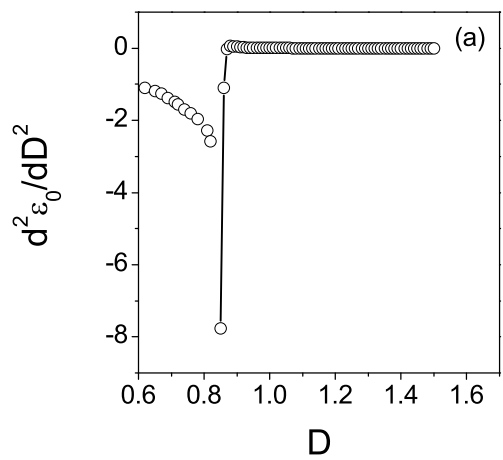
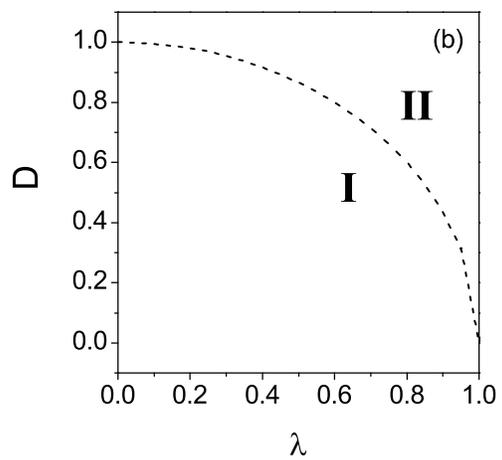
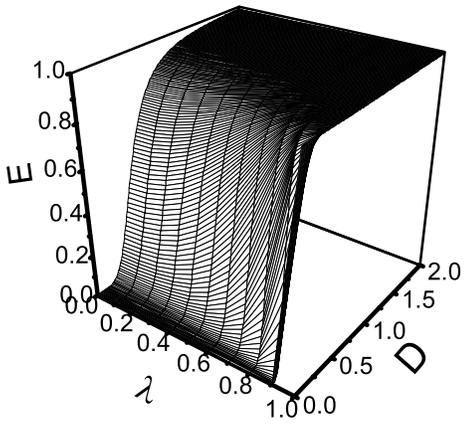
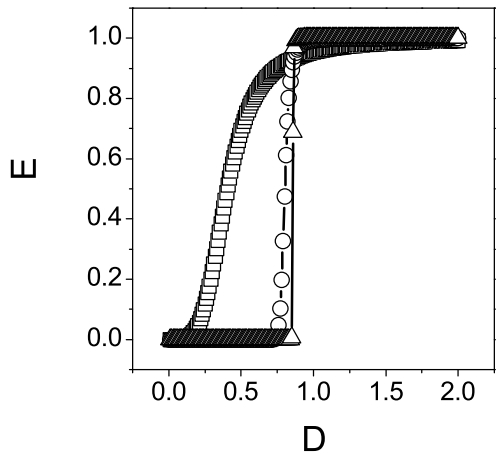
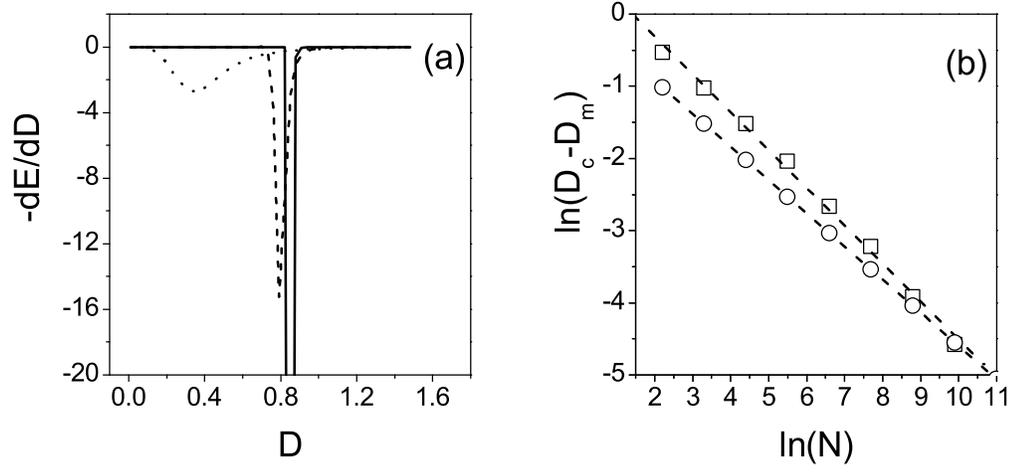
\end{document}